\documentclass[%
superscriptaddress,
 amsmath,amssymb,
 aps,
prb,
twocolumn
]{revtex4-1}

\usepackage[dvipdfmx]{graphicx}
\usepackage{dcolumn}
\usepackage{bm}
\usepackage[dvipdfmx]{color}

\begin{document}

\preprint{To be published in JPSJ}

\title{Optical Process of Linear Dichroism in Angle-Resolved Core-Level Photoemission Reflecting  Strongly Correlated Anisotropic Orbital Symmetry}

\author{Akira Sekiyama}
\author{Yuina Kanai}
\affiliation{Division of Materials Physics, Graduate School of Engineering Science, Osaka University, Toyonaka, Osaka 560-8531, Japan}
\affiliation{RIKEN SPring-8 Center, Sayo, Hyogo 679-5148, Japan}

\author{Arata Tanaka}
\affiliation{Department of Quantum Matter, ADSM, Hiroshima University, Higashi-Hiroshima, Hiroshima 739-8530, Japan}

\author{Shin Imada}
\affiliation{RIKEN SPring-8 Center, Sayo, Hyogo 679-5148, Japan}
\affiliation{Department of Physical Sciences, Ritsumeikan University, Kusatsu, Shiga 525-8577, Japan}

\date{\today}

\begin{abstract}
We revisit the formulations and simulations of angular distributions in polarization-dependent 
core-level photoemission spectra of strongly correlated electron systems, 
in order to explain the recently discovered 
linear dichroism (LD) in 
the core-level photoemission of 4$f$-based rare-earth compounds. 
Owing to the selection rules for the optical process of core-level excitations, 
the LD originating from the anisotropic outer localized charge distributions 
determined by the ground-state orbital symmetry can be observed. 
Our simulations show that core $d$-level excitations are essential for 
the LD in localized ions having a cubic symmetry, which is absent in the $p$-orbital excitations.
 
\end{abstract}

\maketitle



Ground- and excited-state orbital symmetry or orbital polarization in strongly correlated electron 
systems play crucial roles in their functional properties. 
For example,  the highly two-dimensional characteristics of the conducting carriers in high-temperature 
superconducting cuprates are due to their Cu $3d_{x^2-y^2}$ orbital symmetry 
mixed with the O $2p_{x,y}$ symmetry in the 
CuO$_2$ planes~\cite{Abbate90,SuzukiBSCCOXAS,LSCOXAS}, 
which is split by the effective crystalline electric field (CEF). 
Strongly correlated $4f$-based heavy Fermion compounds show a variety of intriguing phenomena 
such as unconventional 
superconductivity~\cite{CeCu2Si2,MathurN1998,OnukiGJPSJ2007,YbAlB4SN08}, 
multipole ordering~\cite{CeB6Hanzawa84,YatskarPrAg2In,PrPb3JPSJ}, 
and both successive transitions~\cite{OnimaruJPSJRev} 
as a function of temperature. 
The CEF-split ground-state $4f$-orbital symmetry of such compounds is very fundamental 
for examining their properties. 

However, the $4f$-orbital symmetry had not been revealed for realistic materials straightforwardly, 
as seen in an old controversy for CeB$_6$ (Ref.~\onlinecite{CeB6RamanNew84}). 
There are too many adjustable parameters to uniquely determine the symmetry 
by an analysis of inelastic neutron scattering and anisotropy in magnetic susceptibility, 
which has been recognized as a standard experimental technique. 
Inelastic polarized neutron scattering~\cite{WillersCePt3Si} is useful but time-consuming, 
and requires a large single-crystalline sample. 
On the other hand, 
x-ray spectroscopic techniques such as linear dichroism (LD) in soft-X-ray 
absorption (XAS) near rare-earth $M_{4,5}$ 
edges~\cite{WillersCePt3Si,Hansmann08LD,WillersCeTIn5,LDXASYbInNi4,WillersCe122,StrigariLD1210} 
and polarization-dependent non-resonant inelastic x-ray scattering~\cite{NIXSCeCu2Si2}  (NIXS)
have been much improved 
for probing the anisotropic $4f$ charge distributions derived from the orbital symmetry. 

Recently, it has been reported that the ground-state $4f$-orbital symmetry 
can be uniquely determined by linear polarization-dependent angle-resolved core-level photoemission
of strongly correlated rare-earth compounds not only in tetragonal symmetry~\cite{TMori2014,CCGAratani} 
but also in cubic symmetry~\cite{Kanai2015,Hamamoto2017}. 
This technique is promising for revealing the ground- and excited-state orbital symmetry of 
strongly correlated electron systems, as it is complementary to LD-XAS and LD-NIXS. 
It has an advantage that it is applicable to systems having cubic symmetry 
for the LD in angle-resolved core-level photoemission of the system 
with localized anisotropic outer charge distributions. 

In this Letter, we show the formulations and simulations of polarization-dependent angle-resolved 
core-level photoemission, including the so called transition matrix elements, 
which describe the transition probabilities between the initial and final states. 
It can be seen that 
LD is intrinsically observed in the spectra 
for partially filled $4f$ systems even in cubic symmetry 
by the formulations. 
These have been a long-standing pitfall in photoemission spectroscopy of solids. 
The matrix elements in angle-resolved valence-band 
photoemission have been discussed previously~\cite{Gadzuk,Goldberg81,Daimon95,NishimotoG}. 
The photoionization cross-sections and asymmetry parameters 
for the core-level photoemission of ions with 
spherical charge distributions 
have already been established~\cite{Scofield76,Lindau,Pdep1,Pdep2,Pdep3}. 
Nevertheless,  the combined formulations of the core-level photoemission 
for strongly correlated systems with the emission angle- and polarization-dependent 
matrix elements have been lacking to date.    
 
In the core-level photoemission process, one inner-core electron is excited by the incident photon with an energy $h\nu$ from the strongly correlated sites, and this photoelectron with a kinetic energy $E_{\rm K}^{*}$ is detected in its final state.  
The intensity of core-level photoemission spectra of strongly correlated electron systems has so far been expressed using an angle and polarization integrated form~\cite{GSPRB,JoKotani,Imada89La,Imada89Ce},  as a function of $\omega\equiv E_{\rm K}^{*}-h\nu$
\begin{equation}
\rho_{n_c l_c}(\omega)=\sum_{f,m_c,s_c}\left|\langle E_f| a_{\lambda_c} |E_i\rangle \right|^2\delta(\omega + E_f-E_i),
\label{CPES1}
\end{equation} 
where $E_i$ stands for the initial-state energy and $E_f$ denotes the eigenenergy of the final state $f$ with a core hole in a solid. 
Here, the operator $a_{\lambda_c}$ annihilates an electron at the core level with quantum numbers specified using a joint index $\lambda_c\equiv (n_c, l_c, m_c, s_c)$, where $n_c$, $l_c$, $m_c$, and $s_c$ denote the principal, orbital, magnetic, and spin quantum numbers, respectively. 
For the ions with an atomic-like partially filled subshell, the one-electron removal state $a_{\lambda_c}|E_i\rangle$ is not an eigenstate of the final state, 
due to the  Coulomb and exchange interactions between the outer electrons and the created core hole, 
which leads to a multiplet-split multiple-peak structure in the core-level photoemission 
spectra~\cite{Imada89La,Imada89Ce,Sugar1972,Thole1985}. 
Hence, in principle, one can detect the anisotropy of the wave function of the outershell through the interactions between the outer electrons and the core hole 
by measuring the emission angle and light polarization dependence of the spectra. 
However, this possibility has been overlooked until recently.

To deal with LD in the angle-resolved core-level photoemission spectra of a single crystal, we need to start from the form in which the transition matrix elements $M_{\gamma_c}$ are explicitly taken into account, as follows:
\begin{widetext}
\begin{equation}
\rho_{n_c l_c}(\omega,\bm{e},\theta_k,\varphi_k)=\sum_{f,s_c}\left|\sum_{m_c}M_{\gamma_c}\langle E_f | a_{\lambda_c} | E_i\rangle \right|^2\delta(\omega + E_f - E_i ),
\label{CPES2}
\end{equation}
\end{widetext}
where $\bm{e}$ is the unit vector indicating the electric field direction of the incident light, and $\theta_k$ and $\varphi_k$ denote the polar
and azimuthal angles of the observed photoelectrons, respectively. 
Here, another joint index $\gamma_c\equiv(n_c, l_c, m_c)$ [thus $\lambda_c=(\gamma_c,s_c)$] is introduced.
$M_{\gamma_c}$ is represented as 
\begin{equation}
M_{\gamma_c}=\iiint \phi^{*}_{\bm{k}}(\bm{r})(e^{{i}\bm{q}\cdot\bm{r}}\bm{e}\cdot\bm{p})\phi_{\gamma_c}(\bm{r})\,dV,
\label{Mat1}
\end{equation}
where $\phi_{\gamma_c}(\bm{r})$ and $\phi_{\bm{k}}(\bm{r})$ denote the one-electron wave function of the core level with $\gamma_c$ and that of the excited photoelectron with a momentum $\bm{k}$ ($E^{*}_{\rm K }=\hbar^2 k^2/2m$), respectively; $\bm{q}$ is the photon momentum and $\bm{p}=-{i}\hbar\nabla$. 
For simplicity, let us discuss the core-level photoemission process for a single ion in the CEF within the electric dipole transitions, i.e., $\exp({i}\bm{q}\cdot\bm{r})\simeq 1$. 
By using the partial wave expansion~\cite{Gadzuk,Goldberg81,Daimon95,NishimotoG}, the photoelectron wave function excited from the inner core state $\lambda_c$ with the atomic-like one-electron wave function
\begin{equation}
\phi_{\gamma_c}(\bm{r})=R_{n_c l_c}(r)Y_{l_c}^{m_c}(\theta,\varphi)
\label{CoreWF}
\end{equation}
can be expressed as
\begin{widetext}
\begin{equation}
\phi_{\bm{k}}(\bm{r})=4\pi\sum_{l',m'}{i}^{l'}e^{-{i}\delta_{l'}}Y^{m'*}_{l'}(\theta_k,\varphi_k)R_{kl'}(r)Y_{l'}^{m'}(\theta,\varphi),
\label{OneEP} 
\end{equation}
where $\delta_{l'}$ stands for the phase shift and $R_{kl'}(r)$ denotes the radial function of the continuum state. By inserting Eqs.~(\ref{CoreWF}) and (\ref{OneEP}) into Eq.~(\ref{Mat1}), the transition matrix elements are rewritten within the electric dipole approximation as
\begin{equation}
M_{\gamma_c}=4\pi \sum_{l'=l_c\pm 1,m'}(-{i})^{l'}e^{{i}\delta_{l'}}Y_{l'}^{m'}(\theta_k,\varphi_k)P(n_cl_c \to kl')\iint Y^{m'*}_{l'}(\theta,\varphi)(\bm{e}\cdot\bm{\hat{r}})Y_{l_c}^{m_c}(\theta,\varphi)\,d\Omega,
\label{Mat2}
\end{equation}
\begin{equation}
P(n_cl_c \to kl')\propto \int R_{kl'}(r)R_{n_c l_c}(r)r^3\,dr, 
\label{radial}
\end{equation}
where $d\Omega=\sin\theta\,d\theta\,d\phi$ and $\bm{\hat{r}}$ is the unit radial vector. 
We further assume that the $l_c\to l_c+1$ transitions are predominant over the $l_c\to l_c-1$ transitions~\cite{Goldberg81,XTLS}, and therefore the interference effects between the outgoing $ l_c+1$ and $l_c-1$ photoelectron waves are negligible~\cite{MLDFe2pPES}. 
Indeed, with this assumption, the experimental polarization-dependent angle-resolved core-level photoemission spectra have been well reproduced by spectral simulations~\cite{TMori2014,CCGAratani,Kanai2015,Hamamoto2017}. By substituting Eq.~(\ref{Mat2}) into Eq.~(\ref{CPES2}) and omitting the $l_c\to l_c-1$ transitions, we finally obtain
\begin{equation}
\rho_{n_c l_c}(\omega,\bm{e},\theta_k,\varphi_k)\propto \sum_{f,s_c}\left|\sum_{m',m_c}Y_{l_c+1}^{m'}(\theta_k,\varphi_k)A_{l_cm_c}^{m'}(\bm{e})\langle E_f | a_{\lambda_c} | E_i\rangle \right|^2\delta(\omega + E_f - E_i )
\label{CARPES}
\end{equation}
\end{widetext}
where
\begin{equation}
A_{l_cm_c}^{m'}(\bm{e})=\iint Y^{m'*}_{l_c+1}(\theta,\varphi)(\bm{e}\cdot\bm{\hat{r}})Y_{l_c}^{m_c}(\theta,\varphi)\,d\Omega
\label{AProb}
\end{equation}
Here, the term $P(n_cl_c \to kl')$ in Eq.~(\ref{CARPES}) is omitted since it is independent of $\bm{e}$, $\theta_k$, and $\varphi_k$. 
Comparing Eq.~(\ref{CARPES}) with Eq.~(\ref{CPES1}), one can recognize that the spectral weights of the multiplet-split peaks are modulated from those in the isotropic spectral function by the light polarization $A_{l_cm_c}^{m'}(\bm{e})$ and photoelectron angular $Y_{l_c+1}^{m'}(\theta_k,\varphi_k)$ factors in the polarization-dependent angle-resolved core-level photoemission spectra. 
Therefore, the multiplet line shape can show the polarization and angular dependence in the CEF, 
where the coordination axes for the electrons cannot be arbitrarily chosen.

As an example, we have simulated the polarization-dependent angle-resolved core-level photoemission spectra caused by the $l_c\to l_c+1$ transitions on the basis of Eq.~(\ref{CARPES}) for Yb$^{3+}$ ions in cubic symmetry using the XTLS 9.0 program~\cite{XTLS}. 
In general, the reasons for the appearance of LD in the core-level photoemission are twofold: one is the initial state effects, which originate from the anisotropy in the wave function in the initial state, and the other is the final states effect, which is the deformation of the spectral structure caused by the level splitting due to CEF or anisotropic hybridization of the orbitals on the ions surrounding the excited site. 
In 4$f$ systems, the level splitting due to CEF is small compared to the life-time and experimental broadening width of the peaks in the spectra; 
therefore we can neglect the final state effects. 
This renders the LD in the core-level photoemission spectroscopy at the rare-earth ion site an ideal tool to examine the symmetry of the 4$f$ state in the initial state.

In Stevens formalism~\cite{Stevens}, the eightfold degenerate Yb$^{3+}$ $J=7/2$ state with one 4$f$ hole is split by the CEF in cubic symmetry into two doublets ($\Gamma_6$ and $\Gamma_7$) and one quartet ($\Gamma_8$), 
where the $\Gamma_6$, $\Gamma_7$, and $\Gamma_8$ 4$f$ charge distributions are elongated along the [100], [111], and [110] directions, respectively~\cite{LDXASYbInNi4,Kanai2015}. 
In the simulations of the 4$d$ and $n_cp$ ($n_c=$ 4, 5) core-level photoemission spectra, the Coulomb and exchange interactions between the 4$f$ hole and the core hole, the spin-orbit interaction of the core hole and 4$f$ orbitals, and the CEF splitting of the 4$f$ levels are included. 
All the atomic parameters such as the Coulomb and exchange interactions (Slater integrals) and the spin-orbit coupling constants have been obtained by using Cowan's code~\cite{Cowan} based on the Hartree-Fock approximation. 
The Slater integrals (spin-orbit coupling constants) are reduced to 88\% (98\% for the inner-core orbitals and 100\% for the 4$f$ orbital).

\begin{figure}
\begin{center}
\includegraphics[width=8cm]{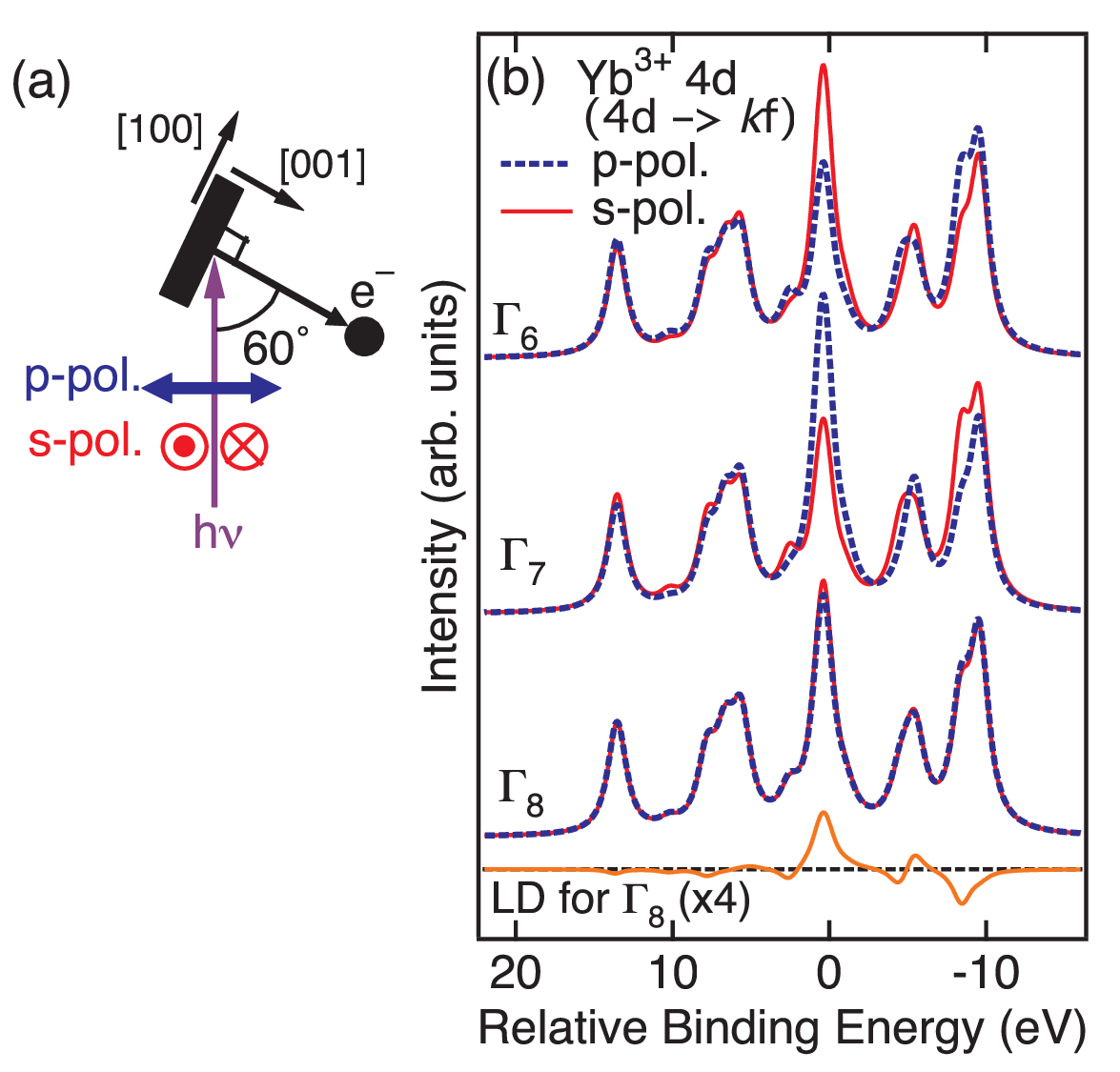}
\end{center}
\begin{center}
\caption{\label{PES4d}(Color online) (a) Geometry of the polarization-dependent 
angle-resolved core-level photoemission. 
(b) Simulated polarization-dependent $4d$ core-level photoemission 
spectra along the [001] direction of Yb$^{3+}$ ions, assuming the CEF-split 
ground state in cubic symmetry. 
The relative binding energy corresponds to $-\omega$. 
The Gaussian and Lorentzian broadenings are set to 0.4 and 1.2 eV, respectively, 
as full widths at half maximum (FWHM).
The spectra are normalized by the overall $4d$ spectral weight. 
LD for the $\Gamma_8$ ground state is also shown. }
\end{center}
\end{figure}

The simulated polarization-dependent 4$d$ core-level photoemission spectra of Yb$^{3+}$ ions 
with the $\Gamma_6$, $\Gamma_7$ and 
$\Gamma_8$ ground states along the [001] direction in the geometry shown in Fig.~\ref{PES4d}(a) are demonstrated in Fig.~\ref{PES4d}(b). 
As theoretically seen in the $3d_{5/2}$ core-level photoemission spectra~\cite{Kanai2015}, 
the LDs defined by the difference in the spectral weight between the $s$- and $p$-polarization configurations as a function of binding energy are also clearly shown in the 4$d$ photoemission spectra. 
For the $\Gamma_6$ and $\Gamma_7$ ground states, the prominent LD is predicted around the relative binding energies of $-$8.5 and 0.4 eV. 
Even for the $\Gamma_8$ ground state, a finite LD is predicted, while the LD is markedly reduced compared with that for the $\Gamma_6$ and $\Gamma_7$ ground states. 
The reduction of LD for the $\Gamma_8$ ground state is ascribed to the fact that the 4$f$ hole spatial distribution is the nearest to a spherical shape among these three states~\cite{Kanai2015}. The simulations suggest that the finite angular dependence of the multiplet-split peak structure can be observed for single crystals even if we measure the spectra using a photoelectron spectrometer with an acceptance angle of  $\pm$several degrees at an unpolarized light excitation with no electric field along the propagation direction of the excitation lights.
It should be noted that the core-level spin-orbit coupling itself is not necessary for LD in the angle-resolved core-level photoemission.

\begin{figure}
\begin{center}
\includegraphics[width=6cm]{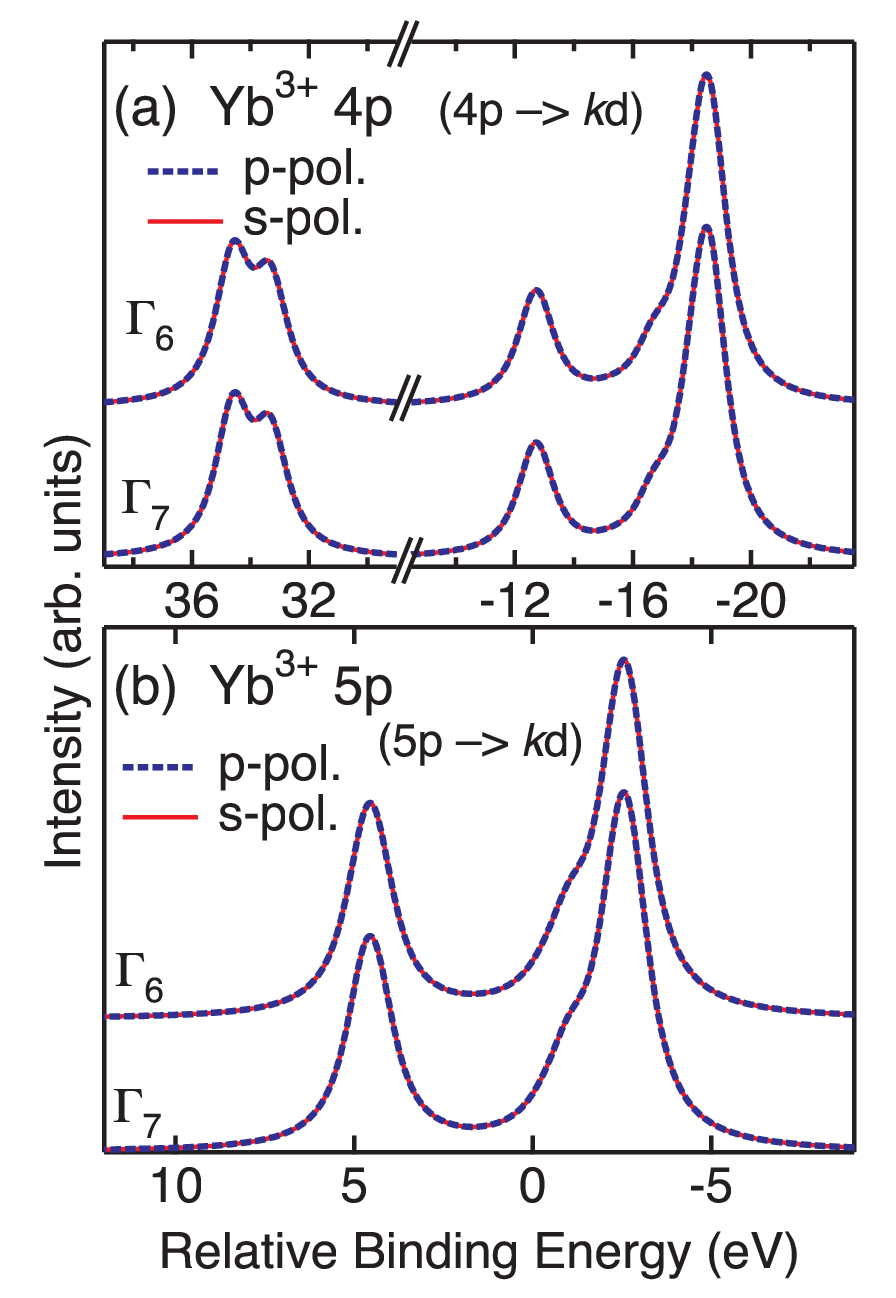}
\end{center}
\begin{center}
\caption{\label{PES4p5p}(Color online) 
(a) Simulated polarization-dependent $4p$ core-level photoemission 
spectra along the [001] direction of Yb$^{3+}$ ions assuming the CEF-split 
$\Gamma_6$ and $\Gamma_7$ ground states in cubic symmetry. 
The geometry of the simulations is the same as that in Fig.~\ref{PES4d}(a). 
(b) Same as (a), but with $5p$ core-level photoemission spectra. 
The Gaussian and Lorentzian broadenings are set to the same as those for the $4d$ 
photoemission spectra in Fig.~\ref{PES4d}(b).}
\end{center}
\end{figure}

Figure~\ref{PES4p5p} shows the simulated angle-resolved 4$p$ and 5$p$ core-level photoemission spectra for the $\Gamma_6$ and $\Gamma_7$ ground states, 
and it can be observed that there is complete absence of LD in the spectra. 
These results are in strong contrast to those in the 4$d$ core-level spectra. 
LD is also absent for the $\Gamma_8$ state, since the isotropic spectra are formed by the sum of the spectra for the $\Gamma_6$, $\Gamma_7$, and  $\Gamma_8$ states weighted by their degeneracy. 
Our simulation results are consistent with the fact that the core-level excitation with $l_c \ge 2$ (irrespective of $n_c$) is essential for observing LD in the angle-resolved photoemission spectra of the systems in cubic symmetry. 
This depends on whether the bases of the $l_c$ core orbital is expressed with a linear combination of the basis sets of multiple irreducible representations, e.g., $d=\Gamma_3\oplus \Gamma_5$ or only one basis set of an irreducible representation, e.g., $p=\Gamma_4$ in the cubic $O$ symmetry. 
For the $n_cp$ core-level spectra, those with the excitation of an electron from any of the core level orbital $p_x$, $p_y$, or $p_z$ or any linear combination of them have the same branching ratio to the final state multiplets, 
and therefore no LD is rigorously concluded, as shown in Fig.~\ref{PES4p5p}. 
On the other hand, for the $4d$ core-level spectra, the branching ratio to the final state multiplets is different depending on whether an electron is excited from those belong to the $\Gamma_3$ representation ($d_{3z^2-r^2}$ and $d_{x^2-y^2}$) or the $\Gamma_5$ representation ($d_{yz}$, $d_{zx}$ and $d_{xy}$), 
and therefore LD is expected.

In summary, we have shown the formulations and simulations of linear polarization-dependent 
angle-resolved core-level photoemission for ions under a CEF. 
A finite LD and angular dependence of the multiplet-split 
core-level spectral shape, which reflects the anisotropic outer localized charge distributions 
determined by the occupied orbital symmetry, are expected 
at the $d$ core-level excitations even for systems having a cubic symmetry. 

We thank Y. Saitoh, H. Fujiwara, and T. Mori for fruitful discussions. 
This work was supported by the Grants-in-Aid for Scientific Research on Innovative Area 
``J-Physics'' (JP16H01074 and JP18H04317), 
a Grant-in-Aid for Scientific Research (JP16H04014) 
from JSPS and MEXT, Japan. 
Y. Kanai was supported by the JSPS Research Fellowships for Young Scientists. 


\references
\bibitem{Abbate90}M. Abbate, M. Sacchi, J. J. Wnuk, L. W. M. Schreurs, Y. S. Wang, R. Lof, and 
J. C. Fuggle, Phys. Rev. B {\bf 42}, 7914 (1990).
\bibitem{SuzukiBSCCOXAS}S. Suzuki, T. Takahashi, T. Kusunoki, T. Morikawa, S. Sato, 
H. Katayama-Yoshida, A. Yamanaka, F. Minami, and S. Takekawa, Phys. Rev. B {\bf 44}, 5381 (1991). 
\bibitem{LSCOXAS}C. T. Chen, L. H. Tjeng, J. Kwo, H. L. Kao, P. Rudolf, F. Sette, and R. M. Fleming, 
Phys. Rev. Lett. {\bf 68}, 2543 (1992). 
\bibitem{CeCu2Si2}F. Steglich, J. Aarts, C. D. Bredl, W. Lieke, D. Meschede, W. Franz, 
and H. Sch\"{a}fer, Phys. Rev. Lett. {\bf 43}, 1892 (1979).
\bibitem{MathurN1998}N. D. Mathur, F. M. Grosche, S. R. Julian, I. R. Walker, 
D. M. Freye, R. K. W. Hselwimmer, and G. G. Lonzarih, Nature (London) {\bf 394}, 39 (1998). 
\bibitem{OnukiGJPSJ2007}R. Settai, T. Takeuchi, and Y. {\=O}nuki, J. Phys. Soc. Jpn. 
{\bf 76}, 051003 (2007), and references therein. 
\bibitem{YbAlB4SN08}S. Nakatsuji, K. Kuga, Y. Machida, T. Tayama, T. Sakakibara, 
Y. Karaki, H. Ishimoto, S. Yonezawa, Y. Maeno, E. Pearson, G. G. Lonzarich, L. Balicas, 
H. Lee, and Z. Fisk, Nat. Phys. {\bf 4}, 603 (2008).
\bibitem{CeB6Hanzawa84}K. Hanzawa and T. Kasuya, J. Phys. Soc. Jpn. {\bf 53}, 1809 (1984). 
\bibitem{YatskarPrAg2In}A. Yatskar, W. P. Beyermann, R. Movshovich, and P. C. Canfield, 
Phys. Rev. Lett. {\bf 77}, 3637 (1996). 
\bibitem{PrPb3JPSJ}D. Aoki, Y. Katayama, R. Settai, Y. Inada, Y. {\=O}nuki, H. Harima, 
and Z. Kletowski, J. Phys. Soc. Jpn. {\bf 66}, 3988 (1997).
\bibitem{OnimaruJPSJRev}T. Onimaru and H. Kusunose, J. Phys. Soc. Jpn. {\bf 85}, 
082002 (2016), and references therein. 
\bibitem{CeB6RamanNew84}E. Zirngiebl, B. Hillebrands, S. Blumenr{\"o}der, G. G{\"u}ntherodt, 
M. Loewenhaupt, J. M. Carpenter, K. Winzer, and Z. Fisk, Phys. Rev. B {\bf 30}, 4052 (1984).
\bibitem{WillersCePt3Si}T. Willers, B. F{\aa}k, N. Hollmann, P. O. K\"{o}rner, Z. Hu, A. Tanaka, 
D. Schmitz, M. Enderle, G. Lapertot, L. H. Tjeng, and A. Severing, 
Phys. Rev. B {\bf 80}, 115106 (2009). 
\bibitem{Hansmann08LD}P. Hansmann, A. Severing, Z. Hu, M. W. Haverkort, C. F. Chang, 
S. Klein, A. Tanaka, H. H. Hsieh, H.-J. Lin, C. T. Chen, B. F{\aa}k, P. Lejay, and L. H. Tjeng, 
Phys. Rev. Lett. {\bf 100}, 066405 (2008).
\bibitem{WillersCeTIn5}T. Willers, Z. Hu, N. Hollmann, P. O. K\"{o}rner, J. Gegner, T. Burnus, 
H. Fujiwara, A. Tanaka, D. Schmitz, H. H. Hsieh, H.-J. Lin, C. T. Chen, E. D. Bauer, 
J. L. Sarrao, E. Goremychkin, M. Koza, L. H. Tjeng, and A. Severing, 
Phys. Rev. B {\bf 81}, 195114 (2010). 
\bibitem{LDXASYbInNi4}T. Willers, J. C. Cezar, N. B. Brookes, Z. Hu, F. Strigari, P. K{\"o}rner, 
N. Hollmann, D. Schmitz, A. Bianchi, Z. Fisk, A. Tanaka, L. H. Tjeng, and A. Severing, 
Phys. Rev. Lett. {\bf 107}, 236402 (2011).
\bibitem{WillersCe122}T. Willers, D. T. Adroja, B. D. Rainford, Z. Hu, N. Hollmann, 
P. O. K\"{o}rner, Y.-Y. Chin, D. Schmitz, H. H. Hsieh, H.-J. Lin, C. T. Chen, E. D. Bauer, 
J. L. Sarrao, K. J. McClellan, D. Byler, C. Geibel, F. Steglich, H. Aoki, P. Lejay, 
A. Tanaka, L. H. Tjeng, and A. Severing, 
Phys. Rev. B {\bf 85}, 035117 (2012). 
\bibitem{StrigariLD1210}F. Strigari, T. Willers, Y. Muro, K. Yutani, T. Takabatake, Z. Hu, Y.-Y. Chin, 
S. Agrestini, H.-J. Lin, C. T. Chen, A. Tanaka, M. W. Haverkort, L. H. Tjeng, and A. Severing, 
Phys. Rev. B {\bf 86}, 081105(R) (2012).
\bibitem{NIXSCeCu2Si2}T. Willers, F. Strigari, N. Hiraoka, Y. Q. Cai, M. W. Haverkort, 
K.-D. Tsuei, Y. F. Liao, S. Seiro, C. Geibel, F. Steglich, L. H. Tjeng, and A. Severing,  
Phys. Rev. Lett. {\bf 109}, 046401 (2012). 
\bibitem{TMori2014}T. Mori, S. Kitayama, Y. Kanai, S. Naimen, H. Fujiwara, A. Higashiya, 
K. Tamasaku, A. Tanaka, K. Terashima, S. Imada, A. Yasui, Y. Saitoh, K. Yamagami, K. Yano, 
T. Matsumoto, T. Kiss, M. Yabashi, T. Ishikawa, S. Suga, Y. {\=O}nuki, T. Ebihara, 
and A. Sekiyama, J. Phys. Soc. Jpn. {\bf 83}, 123702 (2014). 
%
\bibitem{CCGAratani}H. Aratani, Y. Nakatani, H. Fujiwara, M. Kawada Y. Kanai, K. Yamagami, S. Fujioka, S. Hamamoto, K. Kuga, T. Kiss, A. Yamasaki, A. Higashiya, T. Kadono, S. Imada, A. Tanaka, K. Tamasaku, M. Yabashi, T. Ishikawa, A. Yasui, Y. Saitoh, Y. Narumi, K. Kindo, T. Ebihara, and A. Sekiyama, Phys. Rev. B {\bf 98}, 121113(R) (2018).
\bibitem{Kanai2015}Y. Kanai, T. Mori, S. Naimen, K. Yamagami, H. Fujiwara, A. Higashiya, 
T. Kadono, S. Imada, T. Kiss, A. Tanaka, K. Tamasaku, M. Yabashi, T. Ishikawa, F. Iga, 
and A. Sekiyama, J. Phys. Soc. Jpn. {\bf 84}, 073705 (2015). 
%
\bibitem{Hamamoto2017}S. Hamamoto, S. Fujioka, Y. Kanai, K. Yamagami, T. Nakatani, K. Nakagawa, H. Fujiwara, T. Kiss, A. Higashiya, A. Yamasaki, T. Kadono, S. Imada, A. Tanaka, K. Tamasaku, M. Yabashi, T. Ishikawa, K. T. Matsumoto, T. Onimaru, T. Takabatake, and A. Sekiyama, J. Phys. Soc. Jpn. {\bf 86}, 123703 (2017). 
%
\bibitem{Gadzuk}J. W. Gadzuk, Phys. Rev. B {\bf 12}, 5608 (1975). 
\bibitem{Goldberg81}S. M. Goldberg, C. S. Fadley, and S. Kono, 
J. Electron Spectrosc. Relat. Phenom. {\bf 21}, 285 (1981). 
\bibitem{Daimon95}H. Daimon, S. Imada, H. Nishimoto, and S. Suga, 
J. Electron Spectrosc. Relat. Phenom. {\bf 76}, 487 (1995).
\bibitem{NishimotoG}H. Nishimoto, T. Nakatani, T. Matsushita, S. Imada, H. Daimon, 
and S. Suga, J. Phys.: Condens. Matter {\bf 8}, 2715 (1996). 
%
\bibitem{Scofield76}J. H. Scofield, J. Electron Spectrosc. Relat. Phenom. {\bf 8}, 129 (1976). 
\bibitem{Lindau}J. J. Yeh and I. Lindau, At. Data Nucl. Data Tables {\bf 32}, 1 (1985). 
\bibitem{Pdep1}M. B. Trzhaskovskaya, V. I. Nefedov, and V. G. Yarzhemsky, 
At. Data Nucl. Data Tables {\bf 77}, 97 (2001).
\bibitem{Pdep2}M. B. Trzhaskovskaya, V. I. Nefedov, and V. G. Yarzhemsky, 
At. Data Nucl. Data Tables {\bf 82}, 257 (2002).
\bibitem{Pdep3}M. B. Trzhaskovskaya, V. K. Nikulin, V. I. Nefedov, and V. G. Yarzhemsky, 
At. Data Nucl. Data Tables {\bf 92}, 245 (2006).
%
\bibitem{GSPRB}O. Gunnarsson and K. Sch{\"o}nhammer, Phys. Rev. B {\bf 28}, 4315 (1983); 
{\bf 31}, 4815 (1985). 
\bibitem{JoKotani}T. Jo and A. Kotani, J. Phys. Soc. Jpn. {\bf 55}, 2457 (1986).
\bibitem{Imada89La}S. Imada and T. Jo, J. Phys. Soc. Jpn. {\bf 58}, 402 (1989). 
\bibitem{Imada89Ce}S. Imada and T. Jo, J. Phys. Soc. Jpn. {\bf 58}, 2665 (1989). 
\bibitem{Sugar1972}J. Sugar, Phys. Rev. B {\bf 5}, 1785 (1972). 
\bibitem{Thole1985}B. T. Thole, G. van der Laan, J. C. Fuggle, G. A. Sawatzky, R. C, Karnatak, and J.-M. Esteva, Phys. Rev. B {\bf 32}, 5107 (1985).
%
\bibitem{XTLS}A. Tanaka and T. Jo, J. Phys. Soc. Jpn. {\bf 63}, 2788 (1994).
\bibitem{MLDFe2pPES}It has been known that the interference effects are not negligible for a low $E_{\rm K}^*\lesssim$100 eV [see F. U. Hillebrecht, Ch. Roth, H. B. Rose, W. G. Park, E. Kisker, N. A. Cherepkov, Phys. Rev. B {\bf 53}, 12182 (1996)].
%
\bibitem{Stevens}K.W. H. Stevens, Proc. Phys. Soc., Sect. A {\bf 65}, 
209 (1952).
\bibitem{Cowan}R. D. Cowan, {\it The Theory of Atomic Structure and Spectra} 
(University of California Press, Berkeley, CA, 1981).
%

\end{document}